\documentstyle[psfig]{mn} 
\title{A multicolor CCD photometric study of the open clusters NGC~2866, 
Pismis~19, Westerlund~2, ESO96-SC04, NGC~5617
and NGC~6204\thanks{Based on observations obtained with the telescopes of the
South African Astronomical Observatory.}}
\author[G.Carraro and U.Munari]        
{Giovanni Carraro$^1$ and Ulisse Munari$^2$\\
$^1$Dipartimento di Astronomia, Universit\`a di Padova, Vicolo 
dell'Osservatorio 2, I-35122 Padova, Italy {\tt (giovanni.carraro@unipd.it)}\\
$^2$INAF - Osservatorio Astronomico di Padova, Sede di Asiago, I-36012 Asiago
(VI), Italy {\tt (munari@pd.astro.it)}} 
 
\date{\it Submitted: July 2003} 
\pubyear{2002} 

\begin{document} 
 
\maketitle 
\title{A photometric study of six southern open clusters} 
 
\begin{abstract} 
Accurate and deep {\em BVI} CCD photometry is presented for the core regions
of six southern compact open clusters (supplemented in one case by $U$ and
$R$ bands). These clusters generally have only scanty photometric data published,
in some case only photoelectric or photographic. Together with adoption of
modern reference isochrones, this has allowed us to derive improved values for
their distance, reddening, age and main sequence morphology.
\end{abstract} 
 

                             \section{Introduction} 

In this paper we present new CCD multicolor photometry for the compact
southern open clusters NGC~2866, Pismis~19, Westerlund~2, ESO96-SC04,
NGC~5617 and NGC~6204, aiming at deriving updated estimates of their
fundamental parameters. The basic parameters of the clusters are summarized
in Table~1. This study is part of a survey of southern open clusters
conducted by us at the South African Astronomical Observatory (SAAO) in
1992. Other 11 young open clusters observed during the same survey have been
already investigated by Munari \& Carraro (1995, 1996a), Munari et
al.(1998),
Barbon et al.(2000) and Sagar et al.(2001).

All clusters considered in this paper have already received some attention
in the literature. Nonetheless, here and there uncertainties remain because
of limited number of measured members, or bright limiting magnitude, or
absence of profile CCD photometry in crowded areas or uncertain data
calibration. This paper intends to offer high quality photometry of the
program cluster and an homogeneous derivation of basic parameters using the
most recent family of isochrones.

The layout of the paper is as follows. In Sect.~2 we briefly summarize
the present knowledge of the clusters under investigation; Sects.~3 and 4
illstrate the observations and data reduction
strategies, respectively, whereas Sect~5
is devoted to the detailed comparison of our photometry with previous ones.
Sect~6 present our results on a cluster by cluster basis, and, finally,
Sect.~7 briefly summarizes the content of the paper.

                        \section{Previous Investigations}

{\bf NGC~2866.} This is a compact, young and poorly populated cluster. It was
firstly studied by Vogt \& Moffat (1973, hereafter VM73) who obtained $UBV$
photoelectric photometry of the four brightest stars. Later Clari\'a (1979,
hereafter Cl79) provided $UBV$ photoelectric photometry of 23 stars, from
which the cluster turned out to have a reddening $E_{B-V}=0.66\pm0.02$ and a
distance of 2.6$\pm$0.2 kpc. No precise estimates are available for its age,
only the suggestion it is probably too old to be a spiral arm indicator.\\
{\bf Pismis~19.} This is an intermediate age open cluster. CCD photometry
has been obtained by Phelps et al. (1994, hereafter Ph94) in $VI$ and by
Piatti et al. (1998a, hereafter Pt98a) in $BVI$. The latter provide also
integrated spectra. They estimated the reddening in the direction of
Pismis~19 to be $E_{B-V}=1.45\pm0.10$, and a 2.40$\pm$0.88 kpc cluster
distance. They also derived a 1.0$\pm$0.2 Gyr age and (from the integrated
spectra) [Fe/H]=--0.1 metallicity.\\
{\bf Westerlund~2.} This is a very young cluster, embedded in nebulosity.
Vogt \& Moffat (1975, hereafter VM75) firstly obtained $UBV$ photoelectric
photometry for 9 stars in this cluster. Later on Moffat et al. (1991,
hereafter Mf91) got $UBV$ CCD photometry of 220 stars, deriving a distance
of about 8 kpc. The cluster was found to exhibit a high differential
reddening. In addition spectral classification for 7 stars was reported, one
being a new identified WN7 Wolf Rayet star which implies for the cluster a
very young age. Finally, Piatti et al. (1998b, hereafter Pt98b) obtained
integrated spectra, from which they derived an age of 2-3 Myr, and an
absorption $A_V=5$ mag.  Interestingly, they revised the cluster distance,
shortening it to 5.7$\pm$0.3~kpc.\\
{\bf ESO96-SC04.} This is a faint cluster of intermediate age. It has
been observed by Ph94 and Carraro et al. (1995, hereafter Ca95) who both
obtained $BV$ CCD photometry.  The latter derived a distance of 11.8 kpc, a
reddening $E_B-V$=0.75~mag and an age somewhat less than 1 Gyr.

\begin{table*}
\caption{Basic parameters of the program clusters.
Coordinates are for the J2000.0 equinox.}
\begin{tabular}{ccccccccc}
\hline
\multicolumn{1}{c}{name} &
\multicolumn{1}{c}{IAU} &
\multicolumn{1}{c}{OCL} &
\multicolumn{1}{c}{other} &
\multicolumn{1}{c}{Trumpler class} & 
\multicolumn{1}{c}{$\alpha$}  &
\multicolumn{1}{c}{$\delta$}  &
\multicolumn{1}{c}{$l$} &
\multicolumn{1}{c}{$b$} \\
&&&&& ({\em hh:mm:ss}) &
($^{\circ}$~:~$^{\prime}$~:~$^{\prime\prime}$) & $^{\circ}$ & $^{\circ}$
\\
&&&\\
NGC~2866        & C0920-509 & 744 & Pismis~13& I2m:b    
&09:22:07 & -51:06:06 & 273.12 & -0.76\\
Pismis~19       & C1426-607 & 921 &          & I1r:b    
&14:30:42 & -60:59:00 & 314.68 & -0.40\\
Westerlund~2    & C1022-575 & 807 & VdB~95   & IV1~p~n:b
&10:24:02 & -57:45:30 & 284.27 & -0.33\\
ESO~96-SC04     &           &     &          &          
&13:15:09 & -65:55:51 & 305.35 & -3.17\\     
NGC~5617        & C1426-605 & 919 & VdB~159  & I2r      
&14:29:48 & -60:43:00 & 314.68 & -0.11\\
NGC~6204        & C1642-469 & 982 & VdB~196  & I3m:b    
&16:46:08 & -47:00:44 & 338.56 & -1.03\\
\hline
\end{tabular}
\end{table*}

\noindent
{\bf NGC~5617.} Photoelectric photometry for 30 stars has been obtained by
Moffat \& Vogt (1975, hereafter MV75), who derived $E_{B-V}=0.52\pm0.03$,
and a distance of 1.34 kpc from the Sun. Subsequently, the cluster was
studied by Haug (1978, hereafter Ha78), who carried out photoelectric and
photographic photometry, found a similar reddening, and a longer 1.82 kpc
distance. CCD $UBV$ photometry has been reported by Kjeldsen \& Frandsen
(1991, hereafter KF91), who found a smaller reddening $E_{B-V}=0.48\pm0.02$,
a larger distance $2.05\pm0.2$ kpc, and an age of 70 Myr age.\\
{\bf NGC~6204.} This cluster was studied by Moffat \& Vogt (1973, hereafter
MV73), who obtained $E_{B-V}=0.39$, and distance of 0.81 kpc from $UBV$
photometry of 13 stars. Somewhat different results are found by KF91, who
placed the cluster at $1.2\pm0.15$ kpc and suggest a reddening
$E_{B-V}=0.45\pm0.03$ and an age of 0.22 Gyr. More recently, new
photoelectric photometry was carried out by Forbes \& Short (1996, hereafter
FS96), who reported a reddening $E_{B-V}=0.51\pm0.07$, the same KS91
distance and an age of $125\pm25$ Myr.\\

                               \section{Observations} 

The observations were carried out between June 4 and 6, and on July 13,
1992, in the Cousins {\em UBVRI} photometric system, using an
RCA SID53612 thinned CCD detector at the {\it f}/15.8 Cassegrain focus of
the 1.0-m Elizabeth Telescope at the South Africa Astronomical Observatory
(SAAO), Sutherland. A 30 $\mu$m square pixel of the 320 $\times$ 512 CCD
corresponded to 0.388 arcsec and the entire chip covered a field of ~2.1
$\times$ 3.3 arcmin$^2$ on the sky. The readout noise for the system was ~73
electrons pixel$^{-1}$, while the number of electrons per ADU was ~11. The
photometric quality of the sky was monitored using an off-axis photoelectric
photometer that continuously measures a bright, non-variable star located
close to the field imaged on the CCD. The 4 nights were of good photometric
quality with seeing ranging between 1.2 and 2.2 arcsec. 
Table~2 lists the log of our observations. Prior to each
exposure, a 330-ms pre-flash was given to avoid charge transfer problems.
Before that the CCD chip was quickly cleaned 50 times and read once. Every
hour a 33-s pre-flash was acquired to obtain a pre-flash pattern. It was
found to be extremely stable with scatter resulting only from photon
statistics. Flat-field exposures were made of the twilight sky. Dark current
frames were secured in night-time dark conditions.

About 40 E- and F-region standards covering a range in brightness ($7.9 \leq
V \leq 10.3$) as well as in color ($0.01 \leq (V-I) \leq 2.4$) were observed
for calibration purposes. The actual timings of the CCD camera shutter was
monitored by a series of LEP-PSD couples that accurately record the {\it
effective} exposure time to 10ms precision, thus permitting the use of
bright photometric standards from the E and F regions. Typically every 1-1.5
hr, we observed one red and one blue standard star at small airmass and one
red and one blue standard at large airmass in order to have several
independent determinations of the primary and secondary coefficients during
the night, as well as system color equations.  They were found to be quite
constant during the observing run. The excellent photometric quality of the
sky and the small values of airmass at which the observations have been done
ensure the accuracy of the data here presented.
\begin{table} 
\tabcolsep 0.10truecm 
\caption{Journal of observations. $N$ denotes the number of stars
measured in the different passbands.} 
\begin{tabular}{ccccc} 
\hline 
\multicolumn{1}{c}{cluster}    & 
\multicolumn{1}{c}{band}    & 
\multicolumn{1}{c}{exposure}&
\multicolumn{1}{c}{date} & 
\multicolumn{1}{c}{N}         \\ 
      &        & (sec)     & (1992)&\\ 

 NGC~2866             &     &                   &       &    \\ 
                      & $B$ &  60,60,180,900    &  Jun~5& 377\\ 
                      & $V$ &  60,60,90,600     &    '' & 377\\ 
                      & $I$ &  60,60,60,600,600 &    '' & 330\\ 
 Pismis~19            &     &                   &       &    \\ 
                      & $B$ &  900,900          &  Jun~4& 372\\ 
                      & $V$ &  600              &    '' & 372\\ 
                      & $I$ &  600              &    '' & 370\\
                      &     &                   &       &    \\
 Westerlund~2         & $U$ &  965              &  Jun~5&  90\\ 
                      & $B$ &  180,180,600,900  &    '' & 139\\ 
                      & $V$ &  60,180,180,600   &    '' & 351\\ 
                      & $R$ &  60,60,180        &    '' & 282\\
                      & $I$ &  60,60,180,900    &    '' & 281\\
 ESO~96-SC04          &     &                   &       &   \\ 
                      & $B$ &  60,60,1200       &  Jun~6& 890\\ 
                      & $V$ &  60,60,900,900    &    '' & 890\\ 
                      & $I$ &  60,60,900        &    '' & 860\\
 NGC~5617             &     &                   &       &   \\ 
                      & $B$ &  90,140,600       & Jul~13& 100\\ 
                      & $V$ &  30,60,400        &    '' & 140\\ 
                      & $I$ &  60,90,600        &    '' & 140\\
 NGC~6204             &     &                   &       &   \\ 
                      & $B$ &  120,240,400      & Jul~13&  50\\ 
                      & $V$ &  30,60,300        &    '' &  75\\ 
                      & $I$ &  30,180,300       &    '' & 101\\
\hline 
\end{tabular} 
\end{table} 
                           \section{Reduction}  

The data were reduced using the computing facilities available at the
Astronomy Department of the University of Padova, Italy and at the European
Southern Observatory in Garching, Germany, with initial data processing
performed in the usual way via IRAF\footnote{IRAF is distributed by the National
Optical Astronomy Observatories, which is operated by the Association of
Universities for Research in Astronomy, Inc., under contract to the National
Science Foundation.} directly at SAAO in Cape Town. Stellar magnitudes were
obtained by using the DAOPHOT software (Stetson 1987, 1992), and following
processing and conversion of the raw instrumental magnitudes into those of
the standard photometric system were done using procedures outlined by
Stetson (1992). Several stars brighter than $V \approx 10.5$ mag could not
be measured because they saturated even on the shortest cluster and field
regions frames.

\begin{table}
\caption{\label{tab:photcoeff}Average photometric coefficients
obtained during June 4--6 and July 13, 1992.} 
\begin{tabular}{ccccc}
\hline
Filter & Ref. Color & $zp$        & $\gamma$ & $k$ \\
\multicolumn{5}{c}{}\\
\multicolumn{5}{c}{\it June 4-6, 1992}\\
\multicolumn{5}{c}{}\\
 $U$   & $(U-B)$    & 18.11$\pm$0.02 & 0.020$\pm$0.010 &
0.46$\pm$0.02 \\
 $B$   & $(B-V)$    & 20.27$\pm$0.01 & 0.082$\pm$0.015 &
0.27$\pm$0.02 \\
 $V$   & $(B-V)$    & 20.31$\pm$0.01 &-0.025$\pm$0.014 &
0.12$\pm$0.02 \\
 $R$   & $(V-R)$    & 20.56$\pm$0.01 & 0.018$\pm$0.020 &
0.09$\pm$0.02 \\
 $I$   & $(V-I)$    & 19.69$\pm$0.01 & 0.045$\pm$0.015 &
0.06$\pm$0.02 \\
\multicolumn{5}{c}{}\\
\multicolumn{5}{c}{\it July 13, 1992}\\
\multicolumn{5}{c}{}\\
 $B$   & $(B-V)$    & 20.33$\pm$0.01 & 0.002$\pm$0.005 &
0.27$\pm$0.02 \\
 $V$   & $(B-V)$    & 20.11$\pm$0.01 &-0.015$\pm$0.011 &
0.12$\pm$0.02 \\
 $I$   & $(V-I)$    & 17.21$\pm$0.01 & 0.005$\pm$0.015 &
0.06$\pm$0.02 \\
\multicolumn{5}{c}{}\\
\hline
\end{tabular}
\end{table}

\begin{table}
\caption{\label{tab:photerr} Global photometric r.m.s. errors as function
of the magnitude in the given band.}
\begin{tabular}{cccccc}
\hline
mag    &  $\sigma_U$ & $\sigma_B$ & $\sigma_V$ & $\sigma_R$
& $\sigma_I$ \\
 &&&&&\\
  9--11 &  0.04 & 0.03 & 0.03 & 0.03 & 0.03 \\
 11--13 &  0.04 & 0.03 & 0.03 & 0.03 & 0.03\\
 13--15 &  0.04 & 0.03 & 0.03 & 0.03 & 0.03\\
 15--17 &  0.05 & 0.03 & 0.03 & 0.03 & 0.03\\
 17--19 &  0.08 & 0.03 & 0.04 & 0.04 & 0.05\\
 19--20 &    -  & 0.04 & 0.05 & 0.07 & 0.09\\
 20--21 &    -  & 0.07 & 0.09 & 0.15 & 0.22\\
\hline
\end{tabular}
\end{table}

In deriving the color equations for the CCD system and evaluating the
zero-points (zp), we have used nightly values of atmospheric extinction
coefficients. The color equation for the CCD system are derived by
performing aperture photometry on the photometric standards. By fitting
least-squares linear regressions in the observed aperture magnitudes as a
function of the standard photometric indices, we obtained the color equation
listed in Table~3, where $\gamma_i$ is the color term and $k_i$ the
extinction coefficients, $M_i$ and $m_i$ the calibrated and instrumental
magnitudes,
zp$_i$ the zero points and $z$ the airmass:
  \begin{equation}
  M_i = m_i + {\rm zp_i} + \gamma_i(M_i - M_j) - k_i z
  \end{equation} 
We estimated global r.m.s. errors following the expression
given in Patat and Carraro (2001, Eq. A3), and the results
are shown in Table~4.

Complete photometric data for the stars observed in the 6 program clusters
are given in electronic form only. They are made available also
via the WEBDA
database\footnote{http://obswww.unige.ch/webda/navigation.html} maintained
by J.-C. Mermilliod.
Astrometric coordinates to J2000 are
provided for all observed stars, as obtained by astrometric calibration of
recorded frames against local USNO-A2.0 stars. 

                  \section{Comparison with previous photometry} 

All program clusters have some photometry already available in literature,
and a comparison is in order assess the accuracy and evaluate
the agreement of the datasets as well as to compare completeness and photometric band coverage.
The comparisons in this section are always in the sense {\em our} data minus
{\em literature} data.\\
{\bf NGC~2866}. We have observed 377 stars, in {\em BVI}, down to
$V$=22~mag.  Only scanty, photoelectric photometry is available in
literature for this cluster. VM73 have measured 4 stars (in {\em UBV} bands)
brighter than 13~mag, and for the 3 stars in common with us it is:
  \begin{eqnarray}
  \Delta V &=& 0.030 \pm 0.026 \nonumber\\
  \Delta (B-V) &=& 0.021 \pm 0.008 \nonumber
  \end{eqnarray}
Cl79 measured 23 stars brighter than 16~mag, and for the 19 in common with us it
is
  \begin{eqnarray}
  \Delta V &=& 0.058 \pm 0.114 \nonumber\\
  \Delta (B-V) &=& -0.028 \pm 0.039 \nonumber
  \end{eqnarray}
The large dispersion in $\Delta V$ seems caused by field stars contaminating
Cl79 photometry in the core of the cluster, where he used a relatively large
diaphragm (7.6 arcsec).\\

     \begin{figure*}
     \centerline{\psfig{file=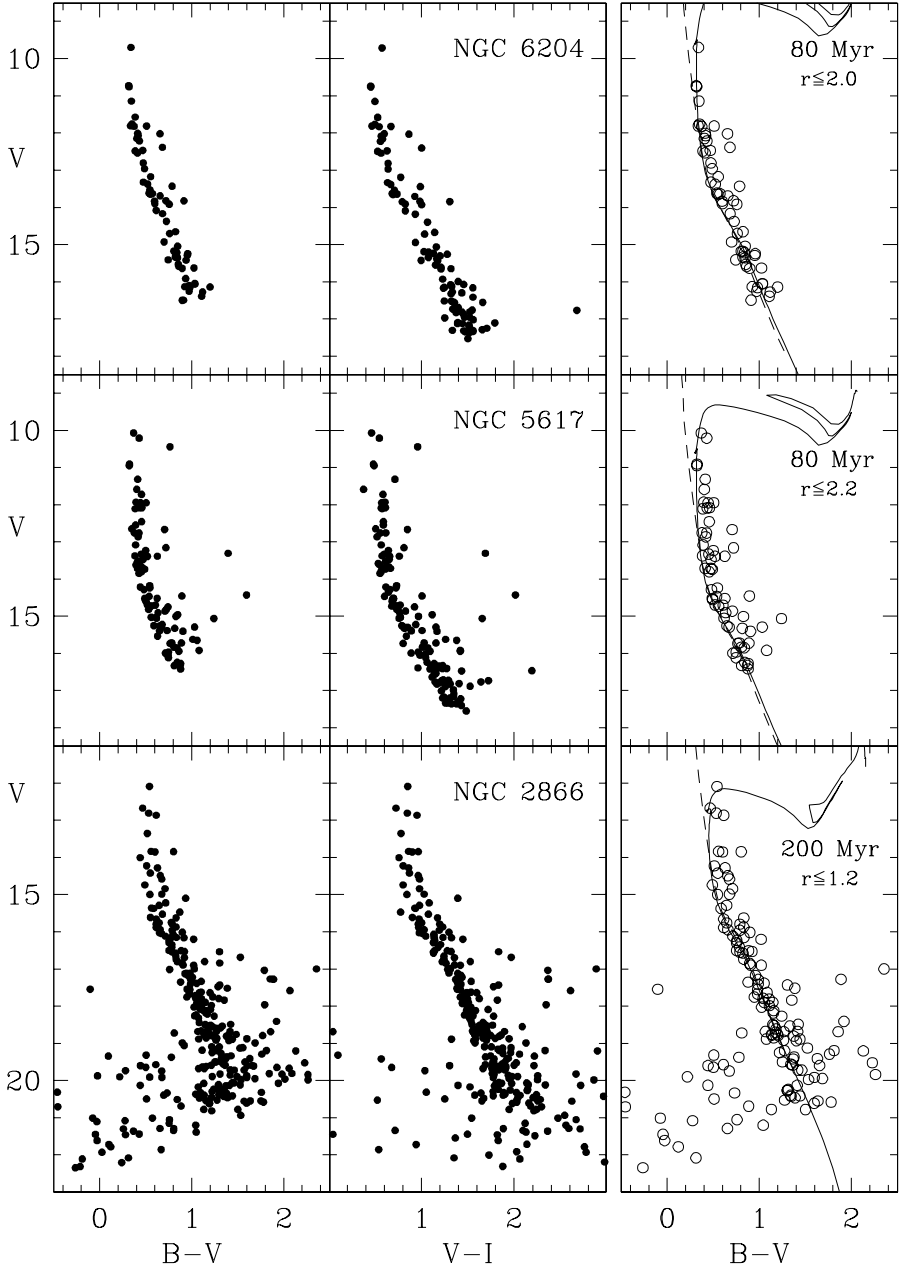,width=12cm}}
     \caption{
     CMDs of the stars in the field of NGC~6204, NGC~5617 and NGC~2866. 
     {\bf Left panels}: all the stars in the $V$ vs ({\em B-V}) diagrams.
     {\bf Central panels}: all the stars in the $V$ vs ({\em V-I}) diagrams.
     {\bf Right panels} : Stars lying within $r$ arcmin from the cluster
     center (indicated for each cluster). The dashed line is the empirical
     ZAMS from Schmidt-Kaler (1982), whereas the solid lines are isochrones
     from Girardi et al.(2000) for the solar metallicity and the indicated age.
     }
     \end{figure*} 
     \begin{figure*}
     \centerline{\psfig{file=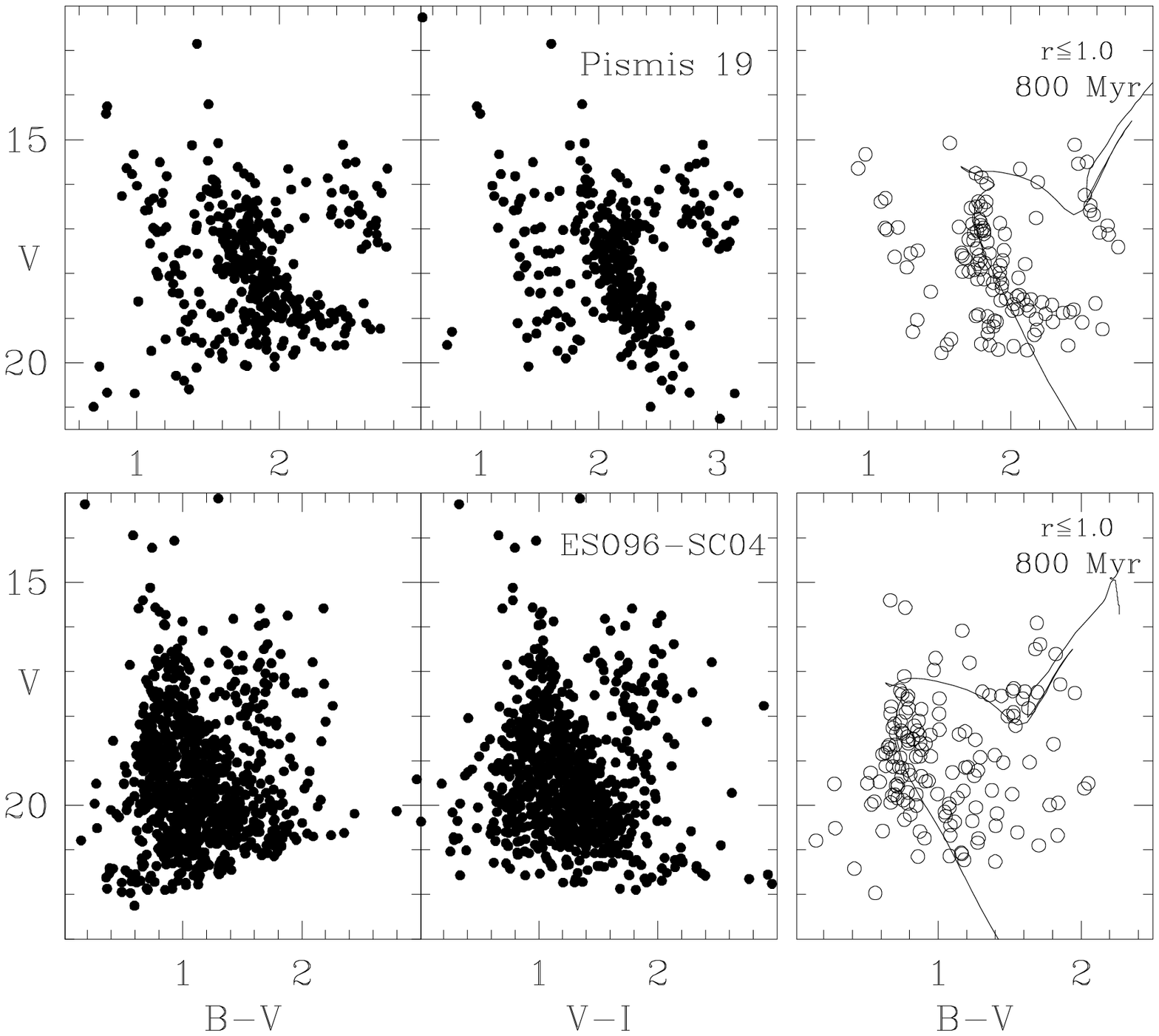,width=12cm}}
     \caption{
     Similar to Figure~1, for the clusters Pismis~19 and ESO96-SC04.
     }
     \end{figure*} 

\noindent
{\bf Pismis~19}. We have observed 372 stars, in {\em BVI}, down to
$V$=21~mag.  {\em BVI} CCD photometry has been presented by Pt98a and {\em
VI} CCD one by Ph94. For the 250 stars in common with Pt98a it is:
  \begin{eqnarray}
  \Delta V &=& ~~0.010 \pm 0.016 \nonumber\\
  \Delta (B-V) &=& ~~0.011 \pm 0.008 \nonumber\\
  \Delta (V-I) &=& -0.009 \pm 0.009 \nonumber
  \end{eqnarray}
and for the 200 in common with Ph94 it is:
  \begin{eqnarray}
  \Delta V &=& -0.041 \pm 0.037 \nonumber\\
  \Delta (V-I) &=& -0.020 \pm 0.087 \nonumber
  \end{eqnarray}
{\bf Westerlund~2}. We have obtained {\em UBVRI} photometry for $\sim$300 stars, down to 19~mag in $V$. Photoelectric photometry has been presented by VM75 for
10 stars and CCD photometry of 89 stars by Mf91, calibrated on the 8 stars
in common with VM75, spanning however a quite limited range in color. Mf91
report that they find insignificant slopes for the color equations, and
therefore they have not color corrected their photometry. Actually, the
color term is not negligible, and comparison of their and our
(color-corrected) $U$ band photometry for the 28 stars in common yields:
  \begin{displaymath}
  \Delta U = 0.076 \times(U-B) - 0.214
  \end{displaymath}
Similar problems were encountered previously (Munari and Carraro 1995) when
comparing our {\em UBVRI} photometry of Bochum~2 with the CCD one by Turbide
and Moffat (1993) that used MV75 photoelectric photometry as local
calibrators. In that case too, as kindly pointed out by A.F.J.Moffat in a
private communication, the CCD photometry was not color corrected. For these
reasons we here refrain from comparison with Mf91 photometry of
Westerlund~2.\\
{\bf ESO~96-SC04}. We have observed 890 stars, in {\em BVI}, down to
$V$=22~mag.  Ph94 presents {\em BVI} CCD observations of this cluster,
which has been used by Ca95 to calibrate
their own {\em BV} photometry. Our {\em BVI} photometry extends $\sim$1 mag
fainter than Ph94, while Ca95 photometry is about $\sim$1 mag deeper than
ours. The comparison with the 600 stars in common with Ph94 yields:
  \begin{eqnarray}
  \Delta V &=& -0.098 \pm 0.049 \nonumber\\
  \Delta (B-V) &=& -0.053 \pm 0.080 \nonumber
  \end{eqnarray}
{\bf NGC~5617}. We have obtained {\em BVI} photometry for 140 stars, down to
17.5~mag in $V$. Ha78 gives photoelectric {\em UBV} photometry of 20 stars,
used to calibrate photographic {\em UBV} photometry of 377 star brighter
than 15~mag in $V$. For the 12 un-blended stars in common with our
photometry, the comparison gives:
  \begin{eqnarray}
  \Delta V &=& 0.069 \pm 0.033 \nonumber\\
  \Delta (B-V) &=& 0.049 \pm 0.031 \nonumber
  \end{eqnarray}
{\em UBV} and Gunn-$r$ CCD photometry is presented by KF91. The comparison
with our photometry for the 59 stars in common gives:
  \begin{eqnarray}
  \Delta V &=& -0.054 \pm 0.027 \nonumber\\
  \Delta (B-V) &=& -0.002 \pm 0.003 \nonumber
  \end{eqnarray}
{\bf NGC~6204}. We have obtained {\em BVI} photometry for $\sim$75 stars,
down to 17.5~mag in $V$. {\em UBV} and Gunn-$r$ CCD photometry is presented
by KF91. The comparison for the 45 stars in common provides:
  \begin{eqnarray}
  \Delta V &=& -0.041 \pm 0.020 \nonumber\\
  \Delta (B-V) &=& -0.003 \pm 0.005 \nonumber
  \end{eqnarray}

The mean deviation of our photometry from the cited references is:\\
  \begin{eqnarray}
  \Delta V &=& -0.009 \pm 0.017 \\
  \Delta (B-V) &=& -0.006 \pm 0.009 
  \end{eqnarray}
which is very small and well within the error of the mean.
This confirms previous findings (cf. Munari \& Carraro
1995, 1996, Munari et al. 1998, Barbon et al. 2000, Sagar et al. 2001) that
the {\em UBVRI} photometry of open clusters that we secured during the 1992
observing runs with the SAAO 1.0m telescope is accurate and well placed on
the international absolute scale.
     \begin{figure*}
     \centerline{\psfig{file=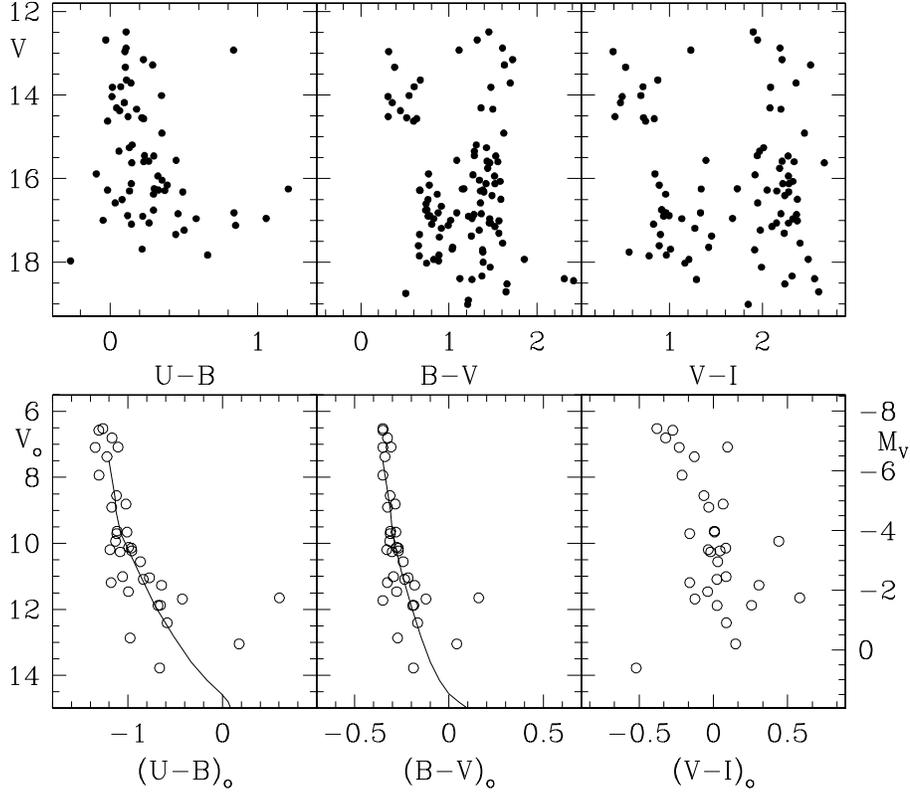,width=12cm}}
     \caption{
     Observed CMD diagrams for all stars in the Westerlund~2 field (top panels) 
     and color-corrected ones
     for the stars represented by open squares in Figure~4 (lower panels), the
differentially
     and highly reddened members of the cluster.
     }
     \end{figure*} 

                    \section{Results}

The main parameters for each cluster
were adopted  the following way:

The reddening of Westerlund~2, for which our photometry includes the
$U$ band, has been derived via the standard $Q-$method. For all the other
clusters,
for which we obtained only {\em BVI} photometry, the reddening has been obtained
from the distribution of member stars on the {\em B-I,  B-V} plane.
As discussed in Munari \& Carraro (1996b, hereafter MC96),
the linear fit to the main sequence on such a plane
\begin{equation}
(B-I) = Q + 2.25 \times (B-V)
\end{equation}
can be expressed in terms of $E_{B-V}$ as
\begin{equation}
E_{B-V} = \frac{Q-0.014}{0.159}
\label{mu2}
\end{equation}
for a standard $R_V=3.1$ extiction law and stars in the 
color range $-0.23 \leq (B-V)_o \leq +1.30$.

Distances and ages of the program clusters have been derived by fitting the
observed CMDs with isochrones from the Padova group (Girardi et al. 2000).
The fit provides an estimate of the reddening too, which has always been very 
close to that obtained with the MC96 method just outlined.
The fit to theoretical isochrones has been performed by eye, paying
particoular attention to the shape of the MS, the position of the brightest
MS stars, the turn-off point and the location of evolved stars, if
present.

The theoretical isochrones are available for a wide range of metallicities.
We have adopted for all clusters those with a solar value because we lacked
firm photometric or spectroscopic determinations of metallicities of
individual clusters, and this information is missing in literature too. The
effect of metallicity has been frequently considered in literature:
increasing it shifts the isochrones fitting toward older ages, larger
distances and smaller reddening.

The results are finally summarized in Table~5, where the basic parameters
are listed together with their uncertainties. The latter correspond to the
shift allowed to isochrone fitting before a mismatch is clearly perceived
by eye inspection.

                    \subsection{NGC~2866}

Cl79 has derived for this cluster $E_{B-V}=0.66\pm0.02$, with no marked
evidence of differential reddening. Our CDMs presented in Figure~1 shows
however a certain breath of the cluster main sequence (MS) suggesting some
$\Delta E_{B-V} \sim$0.1~mag. Applying to our photometry the method by
Munari and Carraro (1996b) to derive the reddening from {\em
BVI} photometry of cluster MS stars, we obtain $E_{B-V}=0.68\pm0.10$. The
result is close to what found by 
Cl79, with a larger error due to our deeper
and wider field photometry that increases the chance of contamination of the
observed cluster ZAMS by field stars.

Cl79, while not providing an estimate of the age, argued that NGC~2866 is
too old to be considered a spiral arm indicator.  From ZAMS and the
isochrone fitting in Figure~1, a distance of 2.6$\pm$3~kpc can be inferred,
in fine agreement with similar findings by Cl79, and ruling out the much
shorter distance of 1.2~kpc proposed by VM73 on the base of photoelectric
photometry of 4 stars. The isochrone fitting indicates an age of 200~Myr.

                       \subsection{Pismis~19}

Pt98a have estimated $E_{B-V}=1.45\pm0.10$, 1.1$\pm0.3$~Gyr age and
2.2$\pm$0.6~kpc distance for this cluster. 

Following the same approach for NGC~2866, our reddening estimate from {\em
BVI} photometry is $E_{B-V}=1.48\pm0.15$, while isochrone fitting in Figure~2
suggests a 0.8~Gyr age and ($m-M$)=15.3$\pm$0.3 distance modulus,
corresponding to a distance of 1.5$\pm$0.4~kpc. The main sequence is well
defined down to the magnitude limit of our photometry, with the turn-off
located at $V \approx 17.0$, $(B-V) \approx 1.80$. The bluer, less populated
sequence visible in the CMDs of Figure~2 is due to foreground galactic disk MS
stars.

The discrepancy in distance and age between us and Pt98a can be -
at least in part - traced to the different set of fitting isochrones
adopted. We have performed a direct fitting of our accurate photometric
data onto a modern family of theoretical isochrones from the Padova group
(Girardi et al. 2000). Pt98a fitting is instead against an set of seven
empirical isochrones built from fiducial points derived from observed CMDs
of 10 clusters calibrated against older theoretical isochrones (including
those of the Padova group released by Bertelli et al. 1994). For such
reasons we are inclined to consider our estimates of distance and age more
accurate than those of Pt98a.

   \begin{figure}
   \centerline{\psfig{file=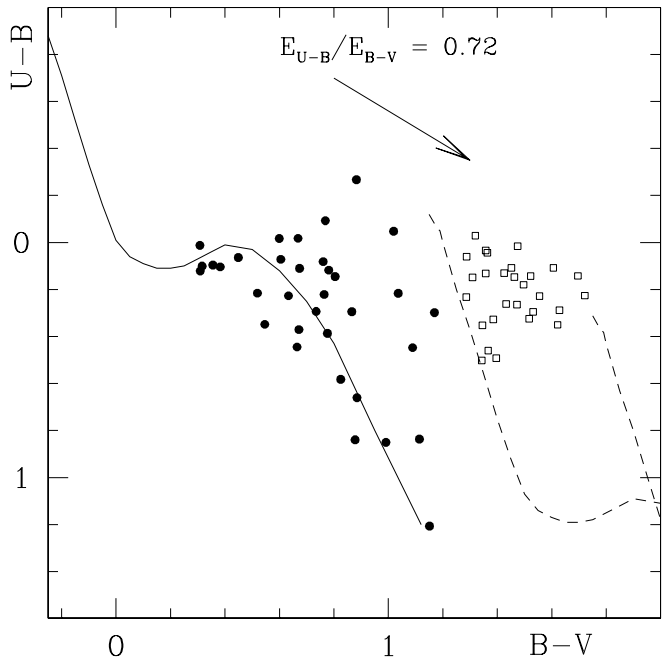,width=8.7cm}}
   \caption{Color-color diagram for the stars in the
   field of Westerlund~2 recorded on $UBV$ frames. The solid line  
   is the empirical ZAMS from Schmidt-Kaler (1982). The
   two dashed lines are the ZAMS shifted by $E_{B-V}$=1.5 and 2.1, to bracket
   the highly and differentially reddened cluster members, represented by
   open squares. Filled circles mark foreground galactic disk stars.
   }
   \end{figure} 

                      \subsection{Westerlund~2}

This is a rather compact open cluster believed to be responsible for the
excitation of the core the HII region RCW~49 (Belloni \& Mereghetti 1994).
Mf91 and Shara et al. (1991) report the discovery of a WN7
Wolf-Rayet member, indicating that Westerlund~2 is very young.

The observed CMDss for Westerlund 2 are presented in
Figure~3. They look very sparse and depicting two ``sequences''. When the
observed stars are plotted in a {\em U--B/B--V} diagram (cf. Figure~4), the
stars of the bluer sequence appear to be galactic disk stars along the line
of sight, while those of the redder sequence are cluster members heavily and
differentially reddened by 1.5$\leq E_{B-V}\leq$2.1. Only the latter are
considered in the lower panels of Figure~3 where they are plotted after
reddening correction following the standard $Q$ method. The mean cluster
reddening turns out to be $E_{B-V}=1.67\pm0.23$, in close agreement with
the $A_V \sim$5.0 spectroscopically estimated by Pt98b.

None of the cluster members appear to  clearly evolving away from the
ZAMS.  The isochrone fitting against it is presented in Figure~3, providing a
distance modulus of ($m-M$)=14.0, corresponding to 6.3~kpc. The same
distance is obtained by combining our photometry and the spectral
classification of its brightest members by Mf91. The close placing of all
members on the empirical ZAMS confirms the cluster to be very young, probably
not older than 2~Myr. A similar young age and distance have been estimated by
Pt98b.

                       \subsection{ESO96-SC04}

CMDs of this cluster presented in Figure~2 are evidently largely contaminated
by field stars, as in Ph94 and Ca95 previous CCD photometry of this faint
target.

Application of the MC96 method to {\em BVI} photometry to derive the
reddening provides $E_{B-V}=0.7\pm0.2$ for ESO96-SC04, with the strong
contamination from field stars causing a large uncertainty. Ca95 estimated
a similar $E_{B-V}\sim$0.75.

Our fit to Padova modern isochrones provides and age of 0.8~Gyr and a
12$\pm$1 kpc distance, close to the 0.7~Gyr and 12~kpc derived by Ca96 using
older versions of the Padova isochrones. From radial star counts and radial
behaviour of the observed CMDs, it may be concluded that the whole cluster is
contained within a radius of 1.5 arcmin from its apparent center. The turn-off
appears located at $V \approx 17.5$, $(B-V) \approx 0.90$.

\begin{table}
\caption{Derived fundamental parameters for the program clusters.}
\begin{tabular}{ccccc}
\hline
\multicolumn{1}{c}{name}  &
\multicolumn{1}{c}{$E_{B-V}$} &
\multicolumn{1}{c}{$(m-M)$}  &
\multicolumn{1}{c}{$d$}  &
\multicolumn{1}{c}{age} \\
& (mag) & (mag) & (kpc) & (Myr) \\
&&&&\\
NGC~2866        & 0.66$\pm$0.10 & 14.0$\pm$0.2 &  2.6$\pm$0.3 &
200    \\
Pismis~19       & 1.48$\pm$0.15 & 15.3$\pm$0.3 &  1.5$\pm$0.4 &
800    \\
Westerlund~2    & 1.67$\pm$0.23 & 19.2$\pm$0.4 &  6.4$\pm$0.4 &
$\leq$2\\
ESO~96-SC04     & 0.70$\pm$0.20 & 17.6$\pm$0.6 &    12$\pm$1  &
800    \\     
NGC~5617        & 0.48$\pm$0.05 & 13.0$\pm$0.2 &  2.0$\pm$0.3 &
80     \\
NGC~6204        & 0.46$\pm$0.05 & 12.0$\pm$0.3 &  1.2$\pm$0.2 &
80     \\
\hline
\end{tabular}
\end{table}
                       \subsection{NGC~5617}

To cover enough area of this open cluster (the larger on the sky
among the program
targets), two adjacent fields have been observed, over which cluster members
largely outnumber field stars (cf. Figure 1).

Application of the MC96 method to {\em BVI} photometry to derive the
reddening provides $E_{B-V}=0.48\pm0.05$ for NGC~5617, close to literature
based on photoelectric data (Ha78 has derived $E_{B-V}$=0.53, MV75 found
$E_{B-V}$=0.52$\pm$0.02 and Lindoff 1968 got $E_{B-V}$=0.53$\pm$0.02) and
identical to more recent CCD determinations (KF91 found
$E_{B-V}$=0.48$\pm$0.02).

The main sequence is well defined down to the limit of our photometry, with
$V\leq12.5$ stars evolving away from it. Our fitting to modern isochrones
provides an age of 80~Myr and a distance of 2.0$\pm$0.3 kpc (cf. Figure~1),
close to KF91 values of 70~Myr and 2.05$\pm$0.20 kpc. This is significantly
largen than old values based on photoelectric photometry:
Lindoff (1968) derived 1.1 kpc and MV75 1.3 kpc, while Ha78 got closer with
1.8 kpc.

                       \subsection{NGC~6204}

Application of the MC96 method to {\em BVI} photometry to derive the
reddening provides $E_{B-V}=0.47\pm0.05$ for NGC~6204, which is close to 
and intermediate between KF91 value of $E_{B-V}=0.45\pm0.03$ and FS96
determination of $E_{B-V}=0.51\pm0.07$.

Similarly to NGC~5617, our photometry (cf. Figure~1) is dominated by cluster
members showing a perfectly defined main sequence down to the magnitude
limit, with $V\leq11.5$ stars evolving away from it. The fitting to modern
isochrones provide an age of 80~Myr and a distance of 1.2$\pm$0.2 kpc. In
previous studies, KF91 found 200~Myr and 1.20$\pm$0.15 kpc, while FS96
obtained 125~Myr and 1.2$\pm$0.1 kpc.

Thus the three available CCD investigations of this cluster pretty well
converge on the distance and reddening, with age estimate getting younger
with time, probably reflecting different and improved sets of reference
isochrones.

\section{Conclusions}
We have presented new homogeneous CCD multicolor photometry for
the six Galactic open clusters NGC~2866, Pismis~19, Westerlund~2,
ESO96-SC04, NGC~5617 and NGC~6204. For all of them we have
provided updated estimates of their fundamental parameters useful
in statistical investigations of the open clusters in our Galaxy.
The results are summarized in Table~5, where we report for each cluster
the reddening, the apparent distance modulus, the distance and the age,
together with their uncertainties.

\section*{Acknowledgments} 

The generous telescope time allocation for this project and the financial
support to UM provided by the SAAO is gratefully acknowledged. One of us
(GC) acknowledges financial support from ESO, where part of this work has
been done. 
We finally thank dr. Brian Chaboyer for carefully reading the manuscript.
This study has been financed also by the Italian Ministry of
University, Scientific Research and Technology (MURST) and the Italian Space
Agency (ASI), and made use of Simbad and WEBDA databases.

\end{document}